\documentclass[aps,pra,twocolumn,superscriptaddress,showkeys,amsmath,amssymb,longbibliography]{revtex4-1}		

\usepackage{graphicx}% Include figure files
\usepackage{dcolumn}% Align table columns on decimal point
\usepackage{bm}% bold math
\usepackage[colorlinks,urlcolor=blue,citecolor=blue,linkcolor=blue]{hyperref}% add hypertext capabilities
\usepackage{subfig}
\usepackage[font=small,labelfont=bf,format=plain,justification=centerlast,labelsep=period]{caption}
\usepackage{tikz}
\usetikzlibrary{arrows,shapes,positioning,shadows,svg.path,backgrounds,fit}

\newcommand{\ket}[1]{\left\vert #1 \right>}

\usepackage{verbatim}
\usepackage[normalem]{ulem}
\usepackage{float}
\usepackage{orcidlink}

\begin{document}

\title{Temporal factorization of a non-stationary electromagnetic cavity field}

\author{I. Ramos-Prieto\,\orcidlink{0000-0001-8838-3541}}
\affiliation{Instituto Nacional de Astrof\'isica \'Optica y Electr\'onica\\ Calle Luis Enrique Erro No. 1 Santa Mar\'ia Tonantzintla, Puebla, 72840, Mexico}

\author{R.  Rom\'an-Ancheyta\,\orcidlink{0000-0001-6718-8587}}
\email{ancheyta@fata.unam.mx}
\affiliation{Centro de F\'isica Aplicada y Tecnolog\'ia Avanzada, Universidad Nacional Autónoma de M\'exico, Boulevard Juriquilla 3001, Quer\'etaro 76230, Mexico}

\author{F. Soto-Eguibar\,\orcidlink{0000-0002-6770-8914}}
\affiliation{Instituto Nacional de Astrof\'isica \'Optica y Electr\'onica\\ Calle Luis Enrique Erro No. 1 Santa Mar\'ia Tonantzintla, Puebla, 72840, Mexico}

\author{J. R\'ecamier}
\affiliation{ Instituto de Ciencias F\'isicas, Universidad Nacional Aut\'onoma de M\'exico\\ Apartado Postal 48-3, Cuernavaca, Morelos 62251, Mexico}

\author{H. M. Moya-Cessa\,\orcidlink{0000-0003-1444-0261}}
\affiliation{Instituto Nacional de Astrof\'isica \'Optica y Electr\'onica\\ Calle Luis Enrique Erro No. 1 Santa Mar\'ia Tonantzintla, Puebla, 72840, Mexico}

\begin{abstract}
 \textcolor{black}{
When an electromagnetic field is confined in a cavity of variable length, real photons may be generated from vacuum fluctuations due to highly nonadiabatic boundary conditions. The corresponding effective Hamiltonian is time-dependent and contains infinite intermode interactions. Considering one of the cavity mirrors fixed and the other describing uniform motion (zero acceleration), we show that it is possible to factorize the entire temporal dependency and write its formal solution, i.e., the Hamiltonian becomes a product of a time-dependent function and a time-independent operator. With this factorization, we prove in detail that the photon production is proportional to the Planck factor involving a velocity-dependent effective temperature. This temperature significantly limits photon generation even for ultra-relativistic motion.} The time-dependent unitary transformations we introduce to obtain temporal factorization help establishing connections with the shortcuts to adiabaticity of quantum thermodynamics and with the quantum Arnold transformation.
\end{abstract}

\date{\today}

\maketitle

\section{Introduction}
Elucidating the dynamics of time-dependent quantum systems is not easy, usually because the corresponding Hamiltonian does not commute with itself at different times. There are perturbative solutions, but many are only valid for short periods~\cite{tannor_2008}. The adiabatic approximation is an option for the long term, if the Hamiltonian is slowly time-dependent~\cite{Sakurai,Berry_1984}; if not, one may still resort to the so-called shortcuts to adiabaticity~\cite{Berry_2009,Chen_2010b}. These shortcuts mimic the adiabatic dynamics in a finite time~\cite{delCampo_2013,Deffner_2014,Vacuum_2023}, and they can be used to manipulate quantum systems before decoherence and dissipation become detrimental~\cite{Guery_2019}. 	 

The quantum harmonic oscillator with time-dependent frequency is the first physical system in which one would like to test the solutions and approaches mentioned above~\cite{Muga_2010_JBP}. For instance, shortcuts to adiabaticity have been implemented during the operation of harmonic quantum heat engines~\cite{Abah_2018} and refrigerators~\cite{Abah_PRR_2020}. Furthermore, the time-dependent harmonic oscillator also serves to study the interesting nonadiabatic behavior of the electromagnetic field. We refer to this as the so-called dynamical Casimir effect (DCE)~\cite{Dodonov_2010,Dodonov_2020}, i.e., the generation of real photons out of the vacuum fluctuations due to nonadiabatic changes in the field's boundary conditions~\cite{Nori_RMP_2012}. \textcolor{black}{Contrary to the static Casimir effect or the Lamb shift, the DCE is a direct manifestation of the existence of the vacuum fluctuations of light.}

The DCE\textcolor{black}{, a name introduced by J. Schwinger~\cite{Schwinger},} is a relativistic effect of the second order and was predicted by G. T. Moore in 1970~\cite{Moore_1970} and experimentally confirmed \textcolor{black}{more than forty years later} only within the platform of superconducting quantum circuits~\cite{Wilson_Nature2011,Paraoanu_PNS_2013}. Moore's original quantum theory \textcolor{black}{of light within a variable-length cavity} did not have a Hamiltonian description; however, significant theoretical progress occurred in 1994 when C. K. Law derived an effective multimodal Hamiltonian for the electromagnetic field that captures the essential features of the DCE~\cite{Law_1994}. This effective Hamiltonian is one of the main subjects of study in the present work and consists of quantum harmonic oscillators with time-dependent frequencies and time-dependent intermode interactions. 
\begin{figure}[t] 
\centering
\includegraphics[scale=0.7]{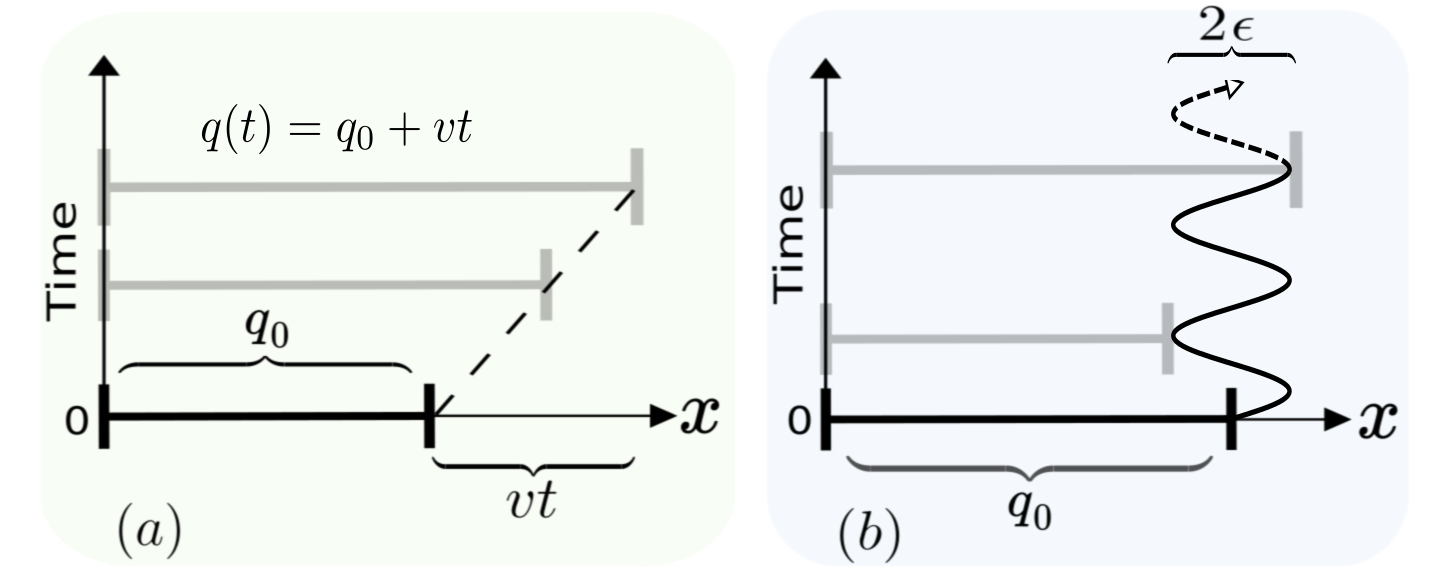}
\caption{\textcolor{black}{Two schematic representations of an ideal one-dimensional non-stationary cavity of initial size $q_0^{}$ (thick horizontal black line). The left mirror is permanently fixed, but the right can move in a given trajectory $x=q(t)$. $(a)$ depicts the uniform motion with $v$ the constant mirror velocity, while $(b)$ is for an oscillatory trajectory (parametric resonance) with a small amplitude modulation $2\epsilon$. The thick gray lines represent the cavity at two different time instants. In the single-mode approximation, the left scenario predicts a minimal generation of photons from the quantum vacuum that is proportional
to the Planck factor involving a velocity-dependent effective temperature; see Eq.~(\ref{eff_temp}). The right one exhibits an expected exponential photon growth; see Fig.~\ref{fig_photons}.}}
\label{dibujo}
\end{figure}

This work shows that under particular circumstances, it is possible to factorize the intricate temporal dependence in the Hamiltonian of an electromagnetic field confined in a non-stationary resonant cavity, see Fig.~\ref{dibujo}. Specifically, when we fix one of the cavity mirrors and the other moves with zero acceleration, we find that a non-trivial time-dependent unitary transformation permits what we dub {\it temporal factorization}\textcolor{black}{, i.e., the system's Hamiltonian becomes a product of a time-dependent function and a time-independent operator.} This factorization enables the resulting Hamiltonian to commute with itself at different times, allowing us to write its formal solution and diagonalize the time-independent part. Our approach makes it easier to determine whether the electromagnetic field gains or loses energy when the cavity contracts or expands, resembling a thermodynamic piston.

\textcolor{black}{We also clarify that although the algebraic structure for the simplest version of the DCE always permits two-photon generation from the vacuum state at arbitrary frequency driving, for the uniformly non-accelerate motion, a Planck factor emerges bounding the photon growth with an effective temperature depending on the mirror’s velocity. This finding contrasts with the Unruh effect~\cite{Unruh_PRD_1976} and related phenomena~\cite{Anatoly_PRL_20118}, where their unbounded  temperatures depend on the proper acceleration~\cite{Davies_1975}.}

We organize our work as follows: Sec.~\ref{Sec2} introduces the non-trivial time-dependent unitary transformations that factorize the temporal dependence of a single quantum harmonic oscillator with arbitrary time-dependent frequency. In Sec.~\ref{temp_fact_1D} we apply such an operational approach to the multi-mode effective Hamiltonian of the non-stationary electromagnetic cavity. Due to the uniform motion of one of the cavity mirrors, temporal factorization is possible. \textcolor{black}{Sec.~\ref{single_mode_case} deals with the single-mode approximation and provides an unreported analytical expression for the photon production. This is proportional to the Planck factor involving a velocity-dependent effective temperature, significantly limiting photon generation even at ultra-relativistic velocities. In Sec.~\ref{two_mode_case}, we provide a diagonalization procedure to show how the interaction (induced by the boundary conditions) between two modes breaks the degeneracy of the cavity spectrum, but without significant effects on the photon production.} Finally, we present our conclusions in Sec.~\ref{Sec4}.

\section{Time-dependent harmonic oscillator}\label{Sec2}
Due to the significance of the harmonic oscillator in quantum physics, and especially for the DCE, we start describing in detail how to solve its time-dependent version using an operator approach. \textcolor{black}{Besides being the preamble of a more challenging problem in the next section, the purpose of studying this model is to introduce the two time-dependent unitary transformations and the corresponding algebraic procedure that factorizes the system's temporal dependence. In addition, we will show how these transformations help establish connections with the shortcuts to adiabaticity of quantum thermodynamics and with the quantum Arnold transformation.} For an arbitrary time-dependent frequency, $\omega(t)$, the oscillator's Hamiltonian is (in this section $\hbar=m=1$) \textcolor{black}{$\hat{\mathcal{H}}_{\texttt{HO}}(t) =\big[\hat{p}^2+\omega^2(t)\hat{x}^2\big]/2$,}
%\begin{equation}\label{H_TDHO}
%\hat{\mathcal{H}}_{\texttt{HO}}(t) =\frac{\hat{p}^2}{2}+\frac{1}{2}{\omega^2(t)}\hat{x}^2,
%\end{equation}
where $\hat{x}$ and $\hat{p}$ are the position and momentum hermitian operators, satisfying $[\hat x, \hat p]={\rm i}$. This oscillator at $t=t_j$ has a frequency $\omega_j\equiv\omega(t_j)$ and energy $\langle \hat{\mathcal{H}}_{\texttt{HO}}(t_j) \rangle$.  Typically, in a given process that starts at $t_0$ and ends at $t_f$, one wants to know how the energies $\langle \hat{\mathcal{H}}_{\texttt{HO}}(t_0) \rangle$ and $\langle \hat{\mathcal{H}}_{\texttt{HO}}(t_f) \rangle$ can be related. This question is of utmost importance in quantum thermodynamics when using the harmonic oscillator as a heat engine. For example, if the process is slow enough, the {adiabatic} limit gives $\langle \hat{\mathcal{H}}_{\texttt{HO}}(t_f) \rangle_{\rm ad}=({\omega_f}/{\omega_0})\langle \hat{\mathcal{H}}_{\texttt{HO}}(t_0) \rangle$~\cite{Jaramillo_2016,delCampo2018}. The answer for an arbitrary driving of $\omega(t)$ is not straightforward; however, with the unitary transformations we introduce to factorize the temporal dependence in the Hamiltonian, this sort of relation appears again if the driving process follows a friction-free trajectory. 
We define the referred time-dependent transformations as~\cite{Moya_2003,Guasti_2003}
\begin{subequations}
\begin{eqnarray}
\hat{\mathcal{D}}_\sigma &=& \exp\Big[\frac{{\rm i}}{2}\frac{\dot{\sigma}(t)}{\sigma(t)}\hat{x}^2\Big], \label{D_transf} \\ 
\hat{\mathcal{S}}_\sigma &=& \exp\Big[-\frac{\rm i}{2}{\ln\sigma(t)}\left(\hat{x}\hat{p}+\hat{p}\hat{x}\right)\Big],\label{S_transf}
\end{eqnarray}
\end{subequations}
where $\sigma(t)$ is, for the moment, an arbitrary well behaved function which we later define and $\dot{\sigma}(t)=d\sigma(t)/dt$. As we see below the notation for $\hat{\mathcal{D}}_\sigma$ and $\hat{\mathcal{S}}_\sigma$ allude to the displacement and squeezing operations. Using the Hadamard lemma~\cite{RossmannW,HallB},   it is not difficult to show the following four transformations
\begin{subequations}
\begin{eqnarray}
\hat{\mathcal{D}}_\sigma^\dagger\hat{p}\hat{\mathcal{D}}_\sigma &=&\hat{p}+\frac{\dot{\sigma}(t)}{\sigma(t)}\hat{x},\qquad 
\hat{\mathcal{D}}_\sigma^\dagger\hat{x}\hat{\mathcal{D}}_\sigma =\hat x, \label{p_transf} \\
\hat{\mathcal{S}}_\sigma^\dagger\hat{p}\hat{\mathcal{S}}_\sigma &=& \frac{1}{\sigma(t)}\hat{p}, \qquad\qquad
\hat{\mathcal{S}}_\sigma^\dagger\hat{x}\hat{\mathcal{S}}_\sigma =\sigma(t)\hat{x}. \label{x_transf}
\end{eqnarray}
\end{subequations}
To make the time-dependent Schrödinger equation ${\rm i}\partial |\Psi(t)\rangle/\partial t=\hat{\mathcal{H}}_{\texttt{HO}}(t)|\Psi(t)\rangle$ invariant under~(\ref{D_transf}), the oscillator's Hamiltonian \textcolor{black}{$\hat{\mathcal{H}}_{\texttt{HO}}(t)$} must transform as~\cite{tannor_2008}
\begin{equation}
\begin{split}
\hat{\mathcal{H}}_\mathcal{D}(t) &= \hat{\mathcal{D}}^\dagger_\sigma\hat{\mathcal{H}}_{\texttt{HO}}(t)\hat{\mathcal{D}}_\sigma-{\rm i}\hat{\mathcal{D}}^\dagger_\sigma\frac{\partial\hat{\mathcal{D}}_\sigma}{\partial t},
\\ 
&=\frac{\hat{p}^2}{2}\!+\!\frac{1}{2}\!\left[\omega^2(t)\!+\!\frac{\ddot{\sigma}(t)}{\sigma(t)}\right]\!\hat{x}^2\!+\!\frac{\dot{\sigma}(t)}{2\sigma(t)}\!\left(\hat{x}\hat{p}\!+\!\hat{p}\hat{x}\right)\!,
\end{split}
\end{equation}
where we have used (\ref{p_transf}), and the state vector changes to $\hat{\mathcal{D}}_\sigma^\dagger|\Psi(t)\rangle$. Applying~(\ref{S_transf}) upon $\hat{\mathcal{H}}_{\mathcal{D}}(t)$ we get
\begin{align}\label{0050}
\hat{\mathcal{H}}_\mathcal{S}(t) &= \hat{\mathcal{S}}^\dagger_\sigma\hat{\mathcal{H}}_\mathcal{D}(t)\hat{\mathcal{S}}_\sigma-{\rm i}\hat{\mathcal{S}}^\dagger_\sigma\frac{\partial\hat{\mathcal{S}}_\sigma}{\partial t},
\nonumber \\ &
=\frac{1}{\sigma^2(t)}\frac{\hat{p}^2}{2}+\frac{1}{2}\left[\omega^2(t)\sigma^2(t)+\ddot{\sigma}(t)\sigma(t)\right]\hat{x}^2. 
\end{align}
For this step the state vector is $\hat{\mathcal{S}}_\sigma^\dagger\hat{\mathcal{D}}_\sigma^\dagger|\Psi(t)\rangle$.

Now the question is, what differential equation should $\sigma(t)$ satisfy to factorize the time dependence in Hamiltonian~(\ref{0050})? This question can be easily answered by recalling that the time-dependent harmonic oscillator admits the Lewis invariant $\hat{I}(t)$. This invariant is a time-dependent operator which can be obtained from $\partial \hat I(t)/\partial t={\rm i}[\hat{I}(t),\hat{\mathcal{H}}_{\texttt{HO}}(t)]$ and it has the following structure~\cite{Lewis_1967} \textcolor{black}{$\hat{I}(t)=\frac{1}{2}{[\rho(t)\hat{p}-\dot{\rho}(t)\hat{x}]^2}+\omega_0^2\hat{x}^2/2\rho^2(t),$}
%\begin{equation}
%\hat{I}(t)=\frac{1}{2}{[\rho(t)\hat{p}-\dot{\rho}(t)\hat{x}]^2}+\frac{1}{2\rho^2(t)}\omega_0^2\hat{x}^2, 
%\end{equation}
where the dimensionless function $\rho(t)$ obeys $\ddot{\rho}(t)+\omega^2(t)\rho(t)=\omega_0^2/\rho^3(t)$.
%\begin{equation}\label{Ermakov_eq}
%\ddot{\rho}(t)+\omega^2(t)\rho(t)=\frac{\omega_0^2}{\rho^3(t)}.
%\end{equation}
This is known as the Ermakov equation, and $\omega_0$ is an arbitrary real constant that we choose to be $\omega(t_0)$. Assuming the  function $\sigma(t)$ as the one that satisfies the Ermakov equation, i.e., $\sigma(t)=\rho(t)$, then by extracting $\ddot\rho(t)\rho(t)$ from the Ermakov equation and substituting it in (\ref{0050}), we get 
\begin{equation}\label{HO_factoriz}
\hat{\mathcal{H}}_\mathcal{S}(t)=\dfrac{1}{\rho^2(t)}\Big(\frac{\hat{p}^2}{2}+\frac{1}{2}\omega_0^2\hat{x}^2\Big).
\end{equation}
Note that the time dependency has been factorized. Due to this factorization, $\hat{\mathcal{H}}_\mathcal{S}(t)=\rho^{-2}(t)\hat{\mathcal{H}}_{\texttt{HO}}(t_0)$ commutes with itself at different times $[\hat{\mathcal{H}}_\mathcal{S}(t_j),\hat{\mathcal{H}}_\mathcal{S}(t_k)]=0$. Therefore, the corresponding time evolution operator can be written as $\exp\big[-{\rm i}\hat{\mathcal{H}}_{\texttt{HO}}(t_0)\int_{t_0}^{t}\rho^{-2}(t')dt'\big]$. To solve the Ermakov differential equation, we must specify the boundary conditions for $\rho(t)$ and \textcolor{black}{$\dot{\rho}(t)$}. In the context of shortcuts to adiabaticity with applications in quantum thermodynamics, especially for the expansion and compression of harmonic \textcolor{black}{ion} traps, one obtains the  boundary conditions by imposing $\hat{I}(t_0)=\hat{\mathcal{H}}_{\texttt{HO}}(t_0)$ and the commutativity between the Lewis invariant and the oscillator Hamiltonian at the final time $t_f$, i.e., $[\hat{I}(t_f),\hat{\mathcal{H}}_{\texttt{HO}}(t_f)]=0$. This requirement can be achieved if $\rho(t_0)=1$, $\dot\rho(t_0)=\ddot\rho(t_0)=0$ and $\rho(t_f)=(\omega_0/\omega_f)^{1/2}$, $\dot\rho(t_f)=\ddot\rho(t_f)=0$. Under such conditions, also known as stationary conditions, $\hat I(t)$ and $\hat{\mathcal{H}}_{\texttt{HO}}(t)$ have simultaneous eigenstates at the beginning and end of a population-preserving evolution. From~(\ref{HO_factoriz}) we note that $\langle \hat{\mathcal{H}}_\mathcal{S}(t_f) \rangle=({\omega_f}/{\omega_0})\langle \hat{\mathcal{H}}_{\texttt{HO}}(t_0) \rangle$, but in contrast with the adiabatic evolution, $t_f-t_0$ is a finite time interval.

With the time evolution operator at hand and the time-dependent unitary transformations, we write the solution for the state vector in the original picture as~\cite{Guasti_2003} \textcolor{black}{$|\Psi(t)\rangle\!=\!\hat{\mathcal{D}}_\rho\hat{\mathcal{S}}_\rho\exp\big[\!-{\rm i}\hat{\mathcal{H}}_{\texttt{HO}}(t_0)\int_{t_0}^{t}\rho^{-2}(t')dt'\big]|\Psi(t_0)\rangle,$}
%\begin{equation}\label{state_vector_ho}
%|\Psi(t)\rangle =
%\hat{\mathcal{D}}_\rho\hat{\mathcal{S}}_\rho
%\exp\left[-{\rm i}\hat{\mathcal{H}}_{\texttt{HO}}(t_0)\int_{t_0}^{t}\rho^{-2}(t')dt'\right]
%|\Psi(t_0)\rangle,
%\end{equation}
where $|\Psi(t_0)\rangle$ is the initial state. In obtaining \textcolor{black}{$|\Psi(t)\rangle$} we used the fact that $\hat{\mathcal{S}}_\rho^\dagger(t_0)=\hat{\mathcal{D}}_\rho^\dagger(t_0)=1$ due to the stationary boundary conditions in the Ermakov equation. As a consequence, an initial Fock state $|n\rangle$ will evolve into a squeezed number state $|\Psi(t_f)\rangle=\exp\big[-{\rm i}E_n\int_{t_0}^{t_f}\rho^{-2}(t')dt'\big]\hat{\mathcal{S}}_\rho(t_f)|n\rangle$~\cite{Oliveira1989}, where $E_n$ are the instantaneous eigenvalue of \textcolor{black}{$\hat{\mathcal{H}}_{\texttt{HO}}(t)$} at $t_0$. Depending on the ratio $\omega_0/\omega_f$, $\rho(t_f)$ is less than or greater than one, which determines the sign of ${\rm ln}\rho(t_f)$ located in the exponent of squeezed transformation~(\ref{S_transf}). Thus, whenever $\omega_0<\omega_f$ ($\omega_0>\omega_f$) the state vector suffers a compression (expansion) in the configuration space, typical of quantum heat engines during the isentropic strokes~\cite{Robnagel_2016,Kosloff_Otto,Abah_2018}. In fact, the variance of the position operator is obtained from (\ref{x_transf}) and gives $(\Delta \hat{x})^2 \propto \rho^2(t_f)=\omega_0/\omega_f$. 

\textcolor{black}{W}e note that if instead of the Ermakov equation, $\sigma$ satisfies the classical harmonic oscillator equation, $\ddot\sigma+\omega^2(t)\sigma=0$~\cite{Ramos_2018a}, the system Hamiltonian also displays temporal factorization. Substituting this equation of motion in~(\ref{0050}), we arrive at the Hamiltonian of the free particle $\hat{\mathcal{H}}_\mathcal{S}(t)=\hat{p}^2/2\sigma^2(t)$. Interestingly, this implies that from a different procedure, we were able to obtain the result given by the quantum Arnold transformation~\cite{Aldaya_2011,Guerrero_2013}. Recall that the Arnold transformation maps the states of the harmonic oscillator into the free particle~\cite{Guerrero_2011}.
%-------------------------------------------------------
\section{Non-stationary electromagnetic cavity field}\label{Sec3}
%\vspace{-1cm}
\textcolor{black}{\subsection{Temporal factorization}\label{temp_fact_1D}}
\textcolor{black}{For many years, the studies of the DCE were performed in the Heisenberg representation, where the quantum electric field operator was built directly from the set of solutions of the bi-dimensional Klein-Gordon wave equation and the corresponding inner product~\cite{Nori_RMP_2012}. However, in this article, we work with Law's effective Hamiltonian of an electromagnetic quantum field in a one-dimensional {empty} cavity with two ideal {(perfectly reflecting)} mirrors; one of them is fixed {at the position $x=0$}, and the other can move in a given trajectory ${x=}q(t)$. Fig.~\ref{dibujo} shows a schematic representation of such a non-stationary cavity with uniform (a) and oscillatory (b) mirror trajectories. In the Lorentz gauge, the classical vector potential, $A(x,t)$, satisfies the wave equation $\partial^2_{x} A(x,t)=c^{-2}\partial^2_{t} A(x,t)$ and the Dirichlet boundary condition $A(0,t)=A(q(t),t)=0$~\cite{DT_Alves_2003}. The procedure to find  Law's Hamiltonian is to expand $A(x,t)$ in terms of the ``instantaneous'' set of mode functions, substitute it in the wave equation, and enforce the boundary condition. Considering the resulting equations of motion as the Euler-Lagrange equations, one can construct a Lagrangian and, using the Legendre transformation, the Hamiltonian; see~\cite{Law_1994,Law_1995} for a detailed derivation. After a standard canonical quantization, the corresponding multi-modal effective Hamiltonian is~\cite{Law_1994}}
\begin{equation}\label{TDHamiltonian}
\hat{H}_{\textrm{eff}}(t)=\sum_{k}\frac{1}{2}\big{\{}\hat{p}_k^2+\omega_k^2(t)\hat{x}_k^2\big{\}}+\frac{\dot{q}(t)}{q(t)}\sum_{\substack{j,k\\j\neq k}}{G}_{kj}^{}\,\hat{p}_k^{}\hat{x}_j^{},
\end{equation}
\textcolor{black}{where $\omega_k^{}(t)=c k\pi/q(t)$ is the instantaneous frequency of the {\textit{k}}-th ($k\in\mathbb{N}$) field mode, $c$ the speed of light, and $\dot q(t)=d q(t)/dt$}. $\hat{x}_j^{}$ and $\hat{p}_k^{}$ are generalized position and momentum operators of the electromagnetic field and satisfy $[\hat{x}_j^{},\hat{p}_{k}^{}]=\mathrm{i}\hbar\,\delta_{jk}^{}$. \textcolor{black}{Note that we will use the usual standard units from now on.} The first term in (\ref{TDHamiltonian}) represents a collection of time-dependent harmonic oscillators, like the one described in Sec.~\ref{Sec2}. The second term describes the characteristic intermode interaction of the DCE with $G_{kj}=(-1)^{k+j}{2kj}/{(j^2-k^2)}$\textcolor{black}{$=\!-G_{jk}$}, determining the \textcolor{black}{antisymmetric coupling coefficient}. For arbitrary mirror trajectories, $\hat{H}_{\rm eff}(t)$ likely does not commute with itself at different times. We are particularly interested in a simple but not trivial trajectory with zero acceleration~\cite{VILLARREAL}. This trajectory is the uniform motion $q(t)=q_0^{}+vt$, where $v$ is a constant and $q_0$ is the \textcolor{black}{right} mirror's initial position\textcolor{black}{, see Fig.~\ref{dibujo}$(a)$}. Depending on the sign of $v$, $q(t)$ causes an expansion or compression of the cavity length. Such a uniform trajectory was realized in the early experiments on laser cavities with moving mirrors at constant velocities~\cite{Smith_1967,Peek_1967,Klinkov_1972}.

Based on the result generated by the time-dependent unitary transformations introduced in the previous section, below we show that the effective Hamiltonian~(\ref{TDHamiltonian}) also displays temporal factorization. Fortunately, due to the parametrization of $q(t)$, we only need to apply the squeezed transformation for each mode. Defining $\hat S(t)=\prod_j\hat{\mathcal{S}}_\sigma^{(j)}$, with $\hat{\mathcal{S}}_\sigma^{(j)}=\exp\left[-{\rm i}\,{\rm ln}\,\sigma(t)(\hat{x}_j\hat{p}_j+\hat{p}_j\hat{x}_j)/2\textcolor{black}{\hbar}\right]$, and $\sigma(t)=\sqrt{q(t)\textcolor{black}{/q_0^{}}}$ \textcolor{black}{is a dimensionless variable such that $\hat{\mathcal{S}}_\sigma^{(j)}|_{t=0}^{}=1$, since $\ln\sigma(0)=0$.} The system Hamiltonian (\ref{TDHamiltonian}) transforms as
$\hat{H}_S(t)=\hat{S}^\dagger(t)\hat{H}_{\textrm{eff}}(t)\hat{S}(t)-\mathrm{i}\hbar\hat{S}^\dagger(t)\partial\hat{S(t)}/\partial t$. Using (\ref{x_transf}) we write the first term as 
\begin{align}
\hat{S}^\dagger(t)\hat{H}_{\textrm{eff}}(t)\hat{S}(t)\
=&\frac{1}{2}\sum_k\biggl\{\frac{\textcolor{black}{q_0^{}}}{q(t)}\hat{p}_k^2+\left[\frac{\textcolor{black}{c}k\pi}{q(t)}\right]^2 \frac{q(t)}{\textcolor{black}{q_0^{}}}\hat{x}^2_k\biggr\}
\nonumber \\ &
+\frac{\dot{q}(t)}{q(t)}\sum_{\substack{j,k\\j\neq k}}G_{kj}\hat{p}_k\hat{x}_j.
\end{align}
The second term is
\begin{equation}
\textrm{i}\hbar\hat{S}^\dagger(t)\frac{\partial\hat{S}(t)}{\partial t}=\frac{1}{4}\frac{\dot{q}(t)}{q(t)}\sum_{j}\left(\hat{x}_j\hat{p}_j+\hat{p}_j\hat{x}_j\right).
\end{equation}
Since $\dot q(t)=v$, the new Hamiltonian simplifies to
\begin{subequations}
\begin{eqnarray}
\hat{H}_S(t)&=&
\frac{\textcolor{black}{q_0^{}}}{q(t)}\bigg[\frac{1}{2}\sum_{k}\bigl\{\hat{p}_k^2+\textcolor{black}{\omega_k^2(0)} \hat{x}_k^2 -\frac{v}{2\textcolor{black}{q_0^{}}}\left(\hat{x}_k\hat{p}_k+\hat{p}_k\hat{x}_k\right)\bigr\}
\label{H_S}\nonumber \\ 
&&\hspace{3cm}+\frac{v}{\textcolor{black}{q_0^{}}}\sum_{\substack{j,k\\j\neq k}}G_{kj}\hat{p}_k\hat{x}_j\bigg],
\\
&\equiv&\textcolor{black}{\frac{q_0^{}}{q(t)}\left[\hat{\mathcal{H}}_S\right],}\label{H_S_b}
\end{eqnarray}
\end{subequations}
where $\hat{\mathcal{H}}_S$ is a time-independent operator, $q(t)=q_0+vt$\textcolor{black}{, and $\omega_k^{}(0)=ck\pi/q_0^{}$}. Note that as in the previous section, the entire time-dependence of $\hat{H}_S(t)$ has been factorized, and now this commutes with itself at different times. \textcolor{black}{S}uch temporal factorization allows us to see that the energy of the system, i.e., $\langle \hat{H}_S(t)\rangle =\langle\hat{\mathcal{H}}_S\rangle\textcolor{black}{q_0^{}}/q(t)$, decreases (increases) when the cavity experiences an expansion (compression) determined by the sign of $v$\textcolor{black}{, resembling a thermodynamic piston.}

On the other hand, it is convenient to define the variable $\tau\equiv v^{-1}\textcolor{black}{q_0^{}}\ln(1+vt/q_0^{})$ such that the Schr\"odinger equation reads $\textrm{i}{\partial\ket{\psi(\tau)}}/{\partial \tau}=\hat{\mathcal{H}}_S\ket{\psi(\tau)}$; its formal solution is $\ket{\psi(t)}=\hat{\mathcal{U}}_S(t)\ket{\psi(0)}$, where $\hat{\mathcal{U}}_S(t)=\exp\big[- {\rm i}\textcolor{black}{q_0^{}}\,v^{-1}{\ln\left(1+vt/q_0^{}\right)}\hat{\mathcal{H}}_S\textcolor{black}{/\hbar} \big]$

\textcolor{black}{
\subsection{Single-mode case}\label{single_mode_case}
}
The above temporal factorization holds for any number of interacting field modes. However, it is known that significant features of the DCE can be obtained assuming that the cavity supports only a single mode. Actually, a three-dimensional non-stationary cavity has a not equidistant spectrum, and its dynamics can be formally reduced to a single one-dimensional parametric oscillator~\cite{Dodonov_PRA_2664,DODONOV1995126}. In the following, we concentrate on such a situation, for instance, considering the principal (lowest) field mode (\ref{H_S}) approximates to
\begin{equation}\label{H_S_approx_1mode}
\hat{H}_S(t)=\frac{q_0^{}}{q(t)}\left[\frac{\hat{p}_1^2}{2}+\frac{1}{2}{\omega_1^2(0)} \hat{x}_1^2 -\frac{v}{4q_0^{}}\left(\hat{x}_1\hat{p}_1+\hat{p}_1\hat{x}_1\right)\right].
\end{equation}
This Hamiltonian represents the simplest version of the DCE, and its set of operators generates the $su(1,1)$ Lie algebra~\cite{Ban_93}: 
\begin{equation}\label{comm_relations_xp}
\begin{split}
    [\hat x_1^2,\hat p_1^2]&=2{\rm i}\hbar (\hat x_1\hat p_1+\hat p_1\hat x_1), \\
    [\hat x_1\hat p_1+\hat p_1\hat x_1,\hat p_1^2] &=4{\rm i}\hbar\, \hat p_1^2,\\
    [\hat x_1\hat p_1+\hat p_1\hat x_1,\hat x_1^2] &=-4{\rm i}\hbar\, \hat x_1^2.
\end{split}
\end{equation}
We show below that (\ref{H_S_approx_1mode}) still contains unreported non-trivial physics concerning the generation of photons from the vacuum state. We can diagonalize it using $\hat{\mathcal D}_v=\exp\big({\rm i}v\hat{x}_1^2/4\hbar q_0^{}\big)$ and the identities (\ref{p_transf}). Such transformation, $\hat{\mathcal D}_v^\dagger\hat{H}_S(t)\hat{\mathcal D}_v$=$\hat{H}_{\rm diag}(t)$, straightforwardly yields
\begin{equation}
\hat{H}_{\rm diag}(t)=\frac{1}{1+{vt}/{q_0^{}}}\left[\frac{\hat{p}_1^2}{2}+\frac{1}{2}\Omega^2_{}(v)\hat{x}_1^2\right],
\end{equation}
where $\Omega(v)\equiv\omega_1^{}(0)\sqrt{1-(v/2\pi c)^2}$ is the time-independent eigenfrequency of the system; when $v\rightarrow 0$,  $\Omega(v)\rightarrow\omega_1^{}(0)$. The time-evolution operator of (\ref{H_S_approx_1mode}) is $\hat{U}_S(t)=\hat{\mathcal{D}}_v\hat{U}_{\rm diag}(t)\hat{\mathcal{D}}_v^\dagger$, where $\hat{U}_{\mathrm{diag}}(t) = \exp\big[-{\rm i}\,f(t)\left(\hat{p}_1^2+\Omega^2_{}(v)\hat{x}_1^2\right)/2\hbar\big]$ and $f(t)=(q_0^{}/v)\ln(1+vt/q_0^{})$.

\textcolor{black}{
To obtain the average photon number, which is one of the most attractive quantities to be studied in the DCE, we need the generalized position and momentum operators in the Heisenberg picture. The Heisenberg picture of an arbitrary time-independent operator, $\hat{O}$, is $\hat{O}(t)=\hat{U}_S^\dagger (t)\hat{O}\,\hat{U}_S(t)$. 
Applying the transformations $\hat{\mathcal{D}}_v$ and $\hat{U}_{\rm diag}(t)$, with the help of (\ref{comm_relations_xp}) we get $\hat{x}_1(t)=\tau_{11}(t)\hat{x}_1+\tau_{12}(t)\hat{p}_1$ and $\hat{p}_1(t) = \tau_{21}(t)\hat{x}_1+\tau_{22}(t)\hat{p}_1$, where
\begin{equation}\label{tau_variables}
\begin{split}
        \tau_{11}(t) &= \cos\left[\Omega(v) f(t)\right]-v\sin\left[\Omega(v)f(t)\right]/2q_0^{}\Omega(v),\\
        \tau_{12}(t) & = \sin\left[\Omega(v) f(t)\right]/\Omega(v),\\
        \tau_{21}(t) &= -\Big[\Omega(v)+\frac{v^2}{4q_0^2\Omega(v)}\Big]\sin\left[\Omega(v) f(t)\right],\\
        \tau_{22}(t) &= \cos\left[\Omega(v)f(t)\right]+v\sin\left[\Omega(v)f(t)\right]/2q_0^{}\Omega(v).
    \end{split}
\end{equation}
For the principal mode ($k=1$), we can unambiguously define the annihilation and creation operators at $t=0$ as:
$\hat{a}=[\omega_1^{}\!(0)\hat{x}_1+{\rm i}\,\hat{p}_1]/\sqrt{2\hbar\omega_1^{}\!(0)}$ and
$\hat{a}^\dagger=[\omega_1^{}\!(0)\hat{x}_1-{\rm i}\,\hat{p}_1]/\sqrt{2\hbar\omega_1^{}\!(0)}$,
such that $[\hat{a},\hat{a}^\dagger]=1$; here, $\hat{a}|0\rangle=0$ defines the vacuum state. The number operator, $\hat{a}^\dagger\hat{a}$, can be extracted~\cite{Sakurai} from $\hat{a}^\dagger\hat{a}=\hat{H}_{\texttt{HO}}^{(1)}\!(0)/\hbar\omega_1^{}\!(0)-1/2$, where $\hat{H}_{\texttt{HO}}^{(1)}\!(0)=[\hat{p}_1^2+\omega_1^2\!(0)\hat{x}_1^2]/2$ is the Hamiltonian of the principal mode at $t=0$, i.e., when the right mirror is in its initial position at $x=q_0^{}$. Using $\hat{x}_1^2(t)$ and $\hat{p}_1^2(t)$, we write $\hat{a}^\dagger\hat{a}$ in the Heisenberg representation. This gives the average number of photons with respect to the vacuum state $\langle\hat a^\dagger \hat a\rangle_0\equiv\langle 0|\hat a^\dagger\!(t)\hat a(t) |0\rangle=\frac{1}{4}[\tau_{11}^2+{\tau_{21}^2}/{\omega_1^2(0)}]+\frac{1}{4}[\tau_{22}^2+\tau_{12}^2\omega_1^2(0)]+(\tau_{12}\tau_{21}-\tau_{11}\tau_{22})/2$, where we have omitted the temporal dependence in the notation of the $\tau_{jk}$ variables. Finally, substituting the explicit values (\ref{tau_variables}) we obtain
}
\textcolor{black}{
\begin{equation}\label{photons_vacuum}
\begin{split}
    \langle\hat a^\dagger \hat a\rangle_0 =
    \frac{1}{\big(\frac{2\pi c}{v}\big)^2\!-\!1}\sin^2\!\left[\frac{1}{2}\ln\Big(1\!+\!\frac{vt}{q_0^{}}\Big)\sqrt{\!\Big(\frac{2\pi c}{v}\Big)^2\!-\!1}\right].
\end{split}
\end{equation}
This is the average number of photons from the quantum vacuum of a single electromagnetic field mode within a non-stationary cavity when one of the cavity mirrors performs a uniform (zero acceleration) motion. As far as we know, (\ref{photons_vacuum}) is an unreported useful analytical expression. Refs.~\cite{Dodonov_PRA_2012, Dodonov_2013, Roman_2017} obtained comparable formulas for parametric quasi-resonant conditions that drastically deviates from the above result. As a function of time, the trigonometric function in (\ref{photons_vacuum}) displays oscillations with values restricted between 0 and 1. That means it is enough to analyze the amplitude term $[(2\pi c/{v})^2\!-\!1]^{-1}$ to find out how much $\langle\hat a^\dagger \hat a\rangle_0$ may grow. The amplitude vanishes when $v\rightarrow 0$, producing no photons, as it should be when the cavity has a fixed length $q(t)\rightarrow q_0^{}$. Remarkably, even for ultra-relativistic mirror velocities, $v\sim c$, photon generation may not be possible since the amplitude term remains very small; its maximum value approximates $1/40= 0.025$. 
}

\begin{figure}[t]
\centering
\includegraphics{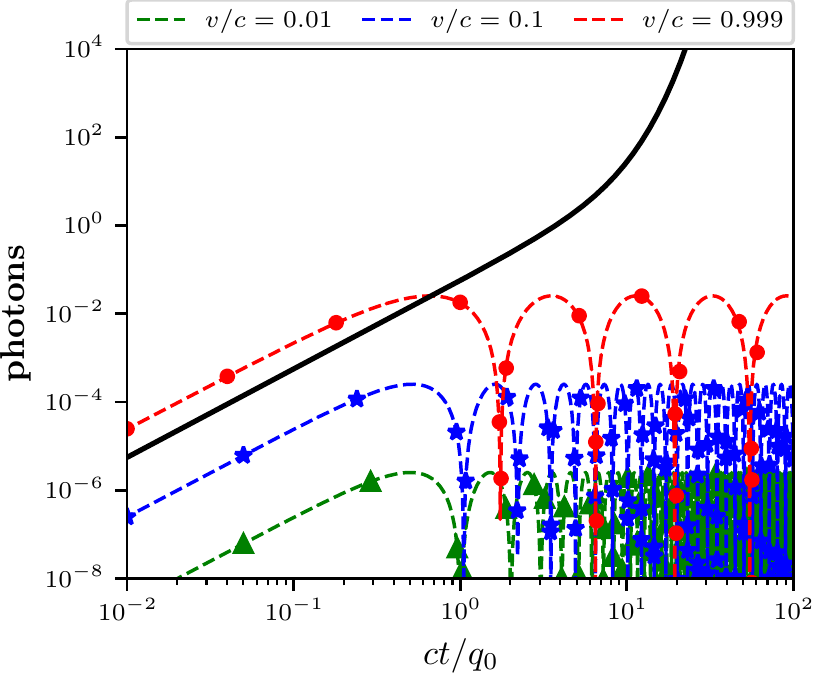}
\caption{\textcolor{black}{Average number of photons from the vacuum state, $\langle \hat a^\dagger \hat a\rangle_0$, as a function of the scaled time, $ct/q_0$, in a single-mode non-stationary electromagnetic cavity. Dashed lines are for the uniform motion [Eq.~(\ref{photons_vacuum})], while the black solid line is for the cavity under parametric resonant conditions, where $\langle \hat a^\dagger \hat a\rangle_0=\sinh^2(\epsilon\pi ct/2q_0^{})$. Photon production for the former is bounded and extremely small for nonrelativistic mirror trajectories $v/c\ll1$. The latter shows the well-known exponential photon growth with $\epsilon=0.15$. Dots, triangles, and stars represent a purely numerical solution using QuTiP~\cite{JOHANSSON20131234}.} }
\label{fig_photons}
\end{figure}

Fig.~\ref{fig_photons} shows, as function of the scaled time $ct/q_0^{}$, the behavior of $\langle\hat a^\dagger \hat a\rangle_0$ for three different values of the ratio $v/c$ (dashed lines). As discussed in the previous paragraph, the number of created photons is bounded and immeasurably small for nonrelativistic mirror trajectories. The dots, triangles, and stars represent a purely numerical solution of $\hat{H}_S(t)$ in (\ref{H_S_approx_1mode}) using QuTiP (Quantum Toolbox in Python)~\cite{JOHANSSON20131234}. For comparison, the black solid line is for the same non-stationary cavity, but under parametric resonant conditions, i.e., the right moving mirror performs an oscillating trajectory with a small modulation amplitude, $\epsilon$, see Fig.~\ref{dibujo}$(b)$. Under these resonant conditions, it is known that the average number of photons grows exponentially as $\langle\hat a^\dagger \hat a\rangle_0=\sinh^2(\epsilon\pi ct/2q_0^{})$~\cite{Dodonov_2020, Dodonov_2010, Nori_RMP_2012}.

\begin{figure}[t]
\begin{center}
\includegraphics{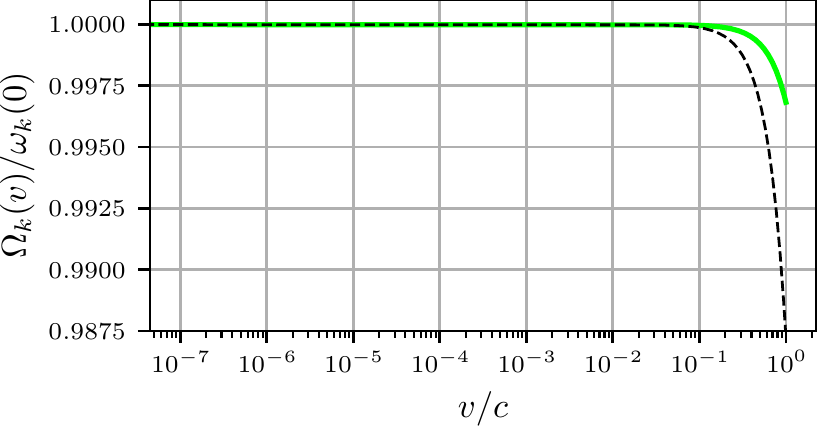}
\caption{\textcolor{black}{Ratio between the eigenfrequecy $\Omega_k(v)$ and the frequency of the static $(t=0)$ $k$ -th cavity mode $\omega_k^{}(0)$. Black (green) dashed (solid) line is for $k=1\,(2)$ and $\Omega_k(v)/\omega_k^{}(0)=\sqrt{1-(v/2k\pi c)^2}$.}}
\label{Omega_v}
\end{center}
\end{figure}

\textcolor{black}{
The amplitude term $[(2\pi c/{v})^2\!-\!1]^{-1}$ can be written as $\omega_1^2(0)\!/\Omega^2(v)\!-\!1$. However, $\Omega(v)$ almost does not change over a large range of $v\!/\!c$ values, see Fig.~\ref{Omega_v}. Thus, $\omega_1^{}\!(0)/\Omega(v)$ is near to one, making the amplitude close to zero; actually, when $v\sim c$, $\Omega(v)\approx 0.98\,\omega_1^{}\!(0)$. Since the system eigenfrequency $\Omega(v)$ does not change significantly, the field will follow approximately an adiabatic dynamic as long as the mirror performs a uniform motion. Therefore, the adiabatic evolution implies an invariant state population without induced transitions and no photon production. Interestingly, we can rewrite (\ref{photons_vacuum}) as $\langle\hat a^\dagger \hat a\rangle_0 =
    \bar{n}_v\, \sin^2\big[\frac{1}{2}\ln(1+vt/q_0^{})\,\bar{n}_v^{-\frac{1}{2}}\big]$, where $ \bar{n}_v=\{\exp[\hbar\,\omega_1^{}\!(0)\!/k_B^{}T_v]-1\}^{-1}$ is the Planck factor involving a velocity-dependent effective temperature
\begin{equation}\label{eff_temp}
T_v=\frac{1}{2}\frac{\hbar\,\omega_1^{}\!(0)}{k_B^{}\!\ln\!\big(2\pi c/v\big)}.
\end{equation}
It is clear that for nonrelativistic motion $T_v\ll 1$, consequently, $\bar{n}_v$ and the photon production vanish. Identifying this temperature from the Planck factor specifically for the uniform motion, which allows temporal factorization, is one of the main contributions of our work. $T_v$ differs substantially from the Unruh temperature, $T_U^{}=\hbar a/(2\pi k_B c)$, a well-known result of quantum field theory~\cite{Unruh_PRD_1976, Davies_1975}. The latter is the effective temperature experienced by a noninertial observer undergoing constant proper acceleration $a$ in the vacuum state. Note that $T_U^{}$ diverges as $a$~\cite{Nori_RMP_2012}; in contrast, $c$ significantly limits $v$ and $T_v$. 
}

\textcolor{black}{
Our results align with previous works showing that uniformly accelerated (including zero acceleration) mirrors do not radiate~\cite{Fulling_1976}. However, surprisingly, our benefit is getting them only using the single-mode version of $\hat{H}_S(t)$, accompanied by a well-known Lie algebra and simple physical explanations. In contrast, \cite{Fulling_1976} needs the energy-momentum tensor and a regularization procedure to give finite results. Furthermore, for any frequency driving, Hamiltonian (\ref{H_S_approx_1mode}) written in terms of $\hat{a}$ and $\hat{a}^\dagger$ is $\hat{H}_S(t)\!=\!\hbar\omega_1^{}\!(t)\hat{a}^\dagger\hat a\!+\!{\rm i}\dot\omega_1^{}\!(t)(\hat{a}^{\dagger 2}\!-\!\hat{a}^2)/4\omega_1^{}\!(t)$~\cite{Law_1994}. Since it contains an explicit squeezing term $(\hat a^{\dagger 2}-\hat a^2)$, this naively suggests a constantly growing photon production, regardless of the mirror trajectory. Here, we contribute to clarifying and demonstrating that such an intuition fails, representing a clear advance over the Law's work~\cite{Law_1994}. 
}

\textcolor{black}{
If instead of the principal cavity mode ($k=1$), we analyze another mode but still within the single-mode approximation, i.e., an uncoupled system, the previous analysis will be the same. The time-independent eigenfrequency of the $k$-th mode now is $\Omega_k^{}(v)=\omega_k^{}(0)\sqrt{1-(v/2k\pi c)}$ and the temperature $T_v(k)=\hbar\omega_k^{}(0)/[2k_B\ln(2 k\pi c/v)]$, meaning that while increasing $k$, $\Omega_k^{}(v)$ will change less, see Fig.~\ref{Omega_v} where $k=2$. Therefore, for the uniform motion, any possible photon production will be much smaller in higher modes than for the principal one.
}

\begin{figure}[t]
\begin{center}
\includegraphics{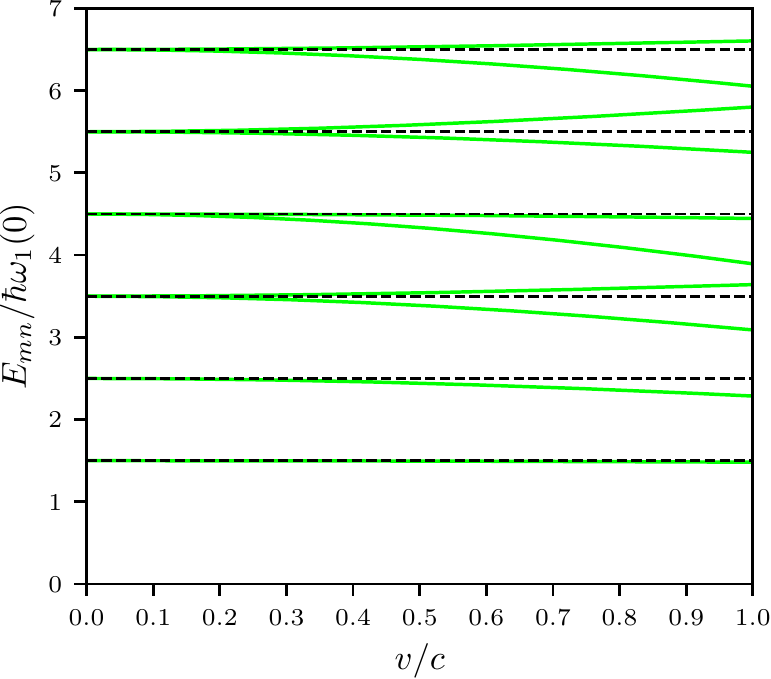}
\caption{\textcolor{black}{Lowest ten eigenvalues $E_{mn}$ of a non-stationary cavity supporting two modes [see Eq.~(\ref{twomodeversion})]. As a function of $v/c$, the uncoupled system (black dashed lines) exhibits eight degenerate states. The intermode interaction induced by the boundary conditions breaks the intrinsic symmetry of the system and, therefore, the degeneracy (green solid lines).}}
\label{Eigenvalues}
\end{center}
\end{figure}

\textcolor{black}{
\subsection{Two-mode case}\label{two_mode_case}
}
\textcolor{black}{In this subsection} we focus on the case where the non-stationary cavity may support two modes. This situation is of particular interest because it is the first that one encounters if one wants to go beyond the single-field mode approximation~\cite{Roman_2017} and deal with the intermode interaction\textcolor{black}{, $G_{kj}$,} induced by the boundary conditions~\cite{Ramos_2021}. The time-independent version of $\hat{\mathcal{H}}_S$ [see (\ref{H_S})] for the two lowest modes is
\begin{equation}\label{twomodeversion}
\begin{split}
\hat{\mathcal{H}}_S&=\sum_{k=1}^2\,\frac{1}{2}\bigl\{\hat{p}_k^2+\textcolor{black}{\omega_k^2(0)} \hat{x}_k^2 -\frac{v}{\textcolor{black}{2q_0^{}}}\left(\hat{x}_k^{}\hat{p}_k^{}+\hat{p}_k^{}\hat{x}_k^{}\right)\bigr\}
\\ 
&\qquad\qquad+{4v}\left(\hat{x}_2\hat{p}_1-\hat{x}_1\hat{p}_2\right)/{3\textcolor{black}{q_0^{}}}.
\end{split}
\end{equation}
\textcolor{black}{The last term of (\ref{twomodeversion}) is the interaction between the field modes 1 and 2. By deliberately ignoring this term and using the results of Sec.~\ref{single_mode_case}, we quickly obtain the energy eigenvalues of the uncoupled Hamiltonian; these are $E_{mn}=\hbar\,\Omega_1^{}\!(v)(m+1/2)+\hbar\,\Omega_2^{}(v)(n+1/2)$, where $m$ and $n$ are non-negative integer numbers. Fig.~\ref{Eigenvalues} shows the lowest ten eigenvalues, see black dashed lines. As a function of $v/c$, we observe only six dashed lines since there are eight degenerate states. Taking the coupling term into account, in appendix~\ref{diagonalization} we describe}, in detail, how (\ref{twomodeversion}) can be diagonalized using three non-trivial unitary transformations, these are $\hat R_1$, $\hat R_2$, and $\hat R_3$ defined in~(\ref{RT}). The corresponding eigenvalues of $\hat{\mathcal{H}}_S$ are $E_{mn}=2\hbar\sqrt{\mu_1\nu_1}(m+1/2)+2\hbar\sqrt{\mu_2\nu_2}(n+1/2)$, where $\mu_j$, $\nu_j$ can be found in~(\ref{muandnu}). Fig.~\ref{Eigenvalues} displays the first ten eigenvalues of (\ref{twomodeversion})\textcolor{black}{; in this case, the intermode interaction breaks the intrinsic symmetry of the system and, therefore, the degeneracy as well (green solid lines). This situation is notorious at ultrarelativistic velocities.}

Diagonalization of $\hat{\mathcal{H}}_S$ in Appendix~\ref{diagonalization} also permits us rewrite $\hat{\mathcal{U}}_S(t)$ in the diagonal basis as
\begin{equation}
\hat{\mathcal{U}}_S(t)=\exp\left[\!-\frac{\mathrm{i}}{\textcolor{black}{\hbar}} \, \textcolor{black}{q_0^{}}v^{-1}{\ln\Big(1+\frac{vt}{q_0^{}}\Big)}\!\sum_{j=1}^2\!\left(\mu_j\hat{p}_j^2+\nu_j\hat{x}_j^2\right)\right],
\end{equation}
while the total time-evolution operator in the original frame is $\hat{U}(t)=\hat{S}(t)\hat{R}_1\hat{R}_2\hat{R}_3\,\hat{\mathcal{U}}_S(t)\hat{R}_3^\dagger\hat{R}_2^\dagger\hat{R}_1^\dagger\hat{S}^\dagger(0).$
\textcolor{black}{With the knowledge of the evolution operator, one can repeat the procedure outlined in Sec.~\ref{single_mode_case} to compute the generation of photons from the vacuum for the two interacting cavity modes. The nine unitary transformations forming $\hat{U}(t)$ make the algebraic approach unwieldy, but the task is doable. It is clear from Fig.~\ref{Eigenvalues} that for $v\ll c$, the energy spectrum coincides with the decoupled system, which, as we discussed in Sec.~\ref{single_mode_case}, has a small photon generation. We performed a purely numerical calculation and confirmed that for $v\sim c$, the photon production from the quantum vacuum in the two-mode interacting case is also minimal.}

To certify the correctness of our   results, we did two verifications:
$a)$ Using the evolution operator, $\hat{U}(t)$, we calculate the operators $\hat{x}_1$, $\hat{x}_2$, $\hat{p}_1$, and $\hat{p}_2$ in the Heisenberg picture and we confirm they satisfy their corresponding Heisenberg equation. 
$b)$
We build a standard linear quantum invariant, $\hat A$, which in the Heisenberg picture reads $\hat{A}(t)=f_1(t) \hat{x}_1+f_2(t) \hat{x}_2+f_3(t) \hat{p}_1+f_4(t) \hat{p}_2$,
where $f_j(t)$ are time-dependent functions, too cumbersome to be shown here, that satisfy the classical Hamilton equations of motion. We prove that $\hat A(t)$ satisfies $d\hat{A}(t)/dt={\rm i}[\hat{\mathcal{H}}_S(t),\hat{A}(t)]+\partial\hat{A}(t)/\partial t=0$; thus, $\hat A(t)$ is indeed invariant when using our solution $\hat U(t)$.\\

\section{Conclusions}\label{Sec4}
Using an operator approach, we factorized the explicit time dependence of the paradigmatic harmonic oscillator with time-dependent frequency and the multimode effective Hamiltonian of a non-stationary cavity field. We dub this result temporal factorization\textcolor{black}{, i.e., the system's Hamiltonian becomes a product of a time-dependent function and a time-independent operator.} For the harmonic oscillator, temporal factorization occurs for any given frequency drive [see Eq.~(\ref{HO_factoriz})], while for the effective Hamiltonian, this is possible when \textcolor{black}{the moving mirror performs a uniform motion (zero acceleration)} [see Eq.~(\ref{H_S_b})]. 

Achieving temporal factorization \textcolor{black}{was helpful as it allowed} their associate Hamiltonian to commute with itself at different times; thus, \textcolor{black}{we obtained} the corresponding evolution operator. \textcolor{black}{It also lets us discern effortlessly how the system gained or lost} energy when undergoing an expansion or compression process, resembling \textcolor{black}{the situation of a thermodynamic piston, especially for the cavity with variable length. Using the oscillator's Lewis invariant,} we got typical results from a quantum thermodynamic cycle's adiabatic (isentropic) stroke. Also, with the time-dependent unitary transformations we use to get factorization, we map the oscillator Hamiltonian into the free particle, making a clear connection with the quantum Arnold transformation.

\textcolor{black}{
For the nonstationary electromagnetic cavity having uniform mirror trajectories, we found that the generation of photons from the vacuum state is proportional to the Planck factor involving a velocity-dependent effective temperature [see Eq.~(\ref{eff_temp})]. This temperature, which differs from the Unruh temperature, strongly limits the photon production in the principal mode, even in the ultra-relativistic limit; the former is much smaller for higher modes. We interpret the low photon growth as a result of an approximate adiabatic dynamic followed by the cavity field's eigenfrequency, mainly for nonrelativistic motion. We went beyond the standard single-mode approximation, and for the two-mode case, we show how the intermode interaction (induced by the boundary conditions) breaks the system degeneracy in the energy levels but with little impact on the photon production. Our results validate the possibility of using the Hamiltonian of the simplest version of the DCE in other related scenarios. For instance, it may serve to evaluate the excitations of an atom near a uniformly moving mirror~\cite{Anatoly_PRR_2019,Anatoly_PRL_20118,Passante_Symmetry_19,Zoller_2010}.}

\textcolor{black}{In principle, temporal factorization would lead us to obtain the spectral response of these non-dissipative time-dependent quantum systems. For instance, we could compute the necessary two-time autocorrelation function of the time-dependent physical spectrum~\cite{Eberly1977} with the resulting evolution operator; for some recent physical examples, see~\cite{roman2019spectral,Salado_Mej_a_2021,Santos_2022}}
Finally, since a free field in curved spacetime is mathematically analogous to a harmonic oscillator with time-dependent frequency~\cite{Jacobson2005},  it would be interesting to explore the implications of our results in the context of quantum field theory in curved spacetimes, including the interaction with an environment~\cite{Ancheyta_2018,Dodonov_1998}. These deserved tasks are far from trivial, and we leave them for \textcolor{black}{further} work.

\textcolor{black}{
\acknowledgments
The authors thank the anonymous referees for their helpful comments and suggestions that significantly improved the paper's content.
J. R. acknowledges partial support from DGAPA-UNAM through project PAPIIT IN109822.}

%%%%%%%%%%%%%%%%%%%%%%%%%%%%%%%%%%%%%%%%%%%%%%%%%%%%%%%%%%%%%%%%%%%%%%%%%%%%%%%%%%%
\appendix

\section{Diagonalization of the system Hamiltonian}\label{diagonalization}
Here we describe in detail how to diagonalize the Hamiltonian $\hat{\mathcal{H}}_S$ in Eq.~(\ref{twomodeversion}) of the main text. Note that this Hamiltonian has a squeezing term in each mode plus the intermode interaction $\propto v(\hat{x}_2\hat{p}_1-\hat{x}_1\hat{p}_2)$ that we must figure out how to get rid of. To diagonalize $\hat{\mathcal{H}}_S$ we introduce three time-independent unitary operators:
\begin{equation}\label{RT}
\begin{split}
\hat{R}_1&=\exp\left[{\rm i}{v}\left(\hat{x}_1^2+\hat{x}_2^2\right)/\textcolor{black}{4\hbar q_0}\right],\\
\hat{R}_2&=\exp\left(-{\rm i}\chi\,\hat{x}_1\hat{x}_2/\textcolor{black}{\hbar}\right),\\
\hat{R}_3&=\exp\left(-{\rm i}\xi\,\hat{p}_1\hat{p}_2/\textcolor{black}{\hbar}\right),
\end{split}
\end{equation}
where $\chi$ and $\xi$ are two arbitrary real parameters that can be later defined. Using the Hadamard lemma \cite{RossmannW,HallB} we can write the following six transformations
\begin{equation}\label{relations}
\begin{split}
\hat{R}_1^\dagger\,\hat{p}_1\hat{R}_1&=\hat{p}_1+\textcolor{black}{\frac{v}{2q_0}}\hat{x}_1,\\
\hat{R}_2^\dagger\,\hat{p}_1\hat{R}_2&=\hat{p}_1-\chi\hat{x}_2,\\
\hat{R}_3^\dagger\,\hat{x}_1\hat{R}_3&=\hat{x}_1+\xi\hat{p}_2,
\end{split}
,\quad
\begin{split}
    \hat{R}_1^\dagger\,\hat{p}_2\hat{R}_1&=\hat{p}_2+\textcolor{black}{\frac{v}{2q_0}}\hat{x}_2,\\
\hat{R}_2^\dagger\,\hat{p}_2\hat{R}_2&=\hat{p}_2-\chi\hat{x}_1,\\
\hat{R}_3^\dagger\,\hat{x}_2\hat{R}_3&=\hat{x}_2+\xi\hat{p}_1.
\end{split}
\end{equation}
We now move to a scenario determined by~(\ref{RT}), i.e., $\hat{R}_3^\dagger\hat{R}_2^\dagger\hat{R}_1^\dagger\hat{\mathcal{H}}_S\hat{R}_1\hat{R}_2\hat{R}_3\equiv\hat{\mathcal{H}}_3$. Evidently, $\hat{\mathcal{H}}_S$ and $\hat{\mathcal{H}}_3$ have the same eigenvalues because $\hat{R}_j$ are unitary operators~\cite{Sakurai}. By taking into account~(\ref{relations}) we get $\hat{\mathcal{H}}_3=\sum_{j=1}^2(\mu_j\hat{p}_j^2+\nu_k\hat{x}_j^2)+\eta_{12}\hat{x}_1\hat{p}_2+\eta_{21}\hat{x}_2\hat{p}_1$,
where the coefficients $\nu_i$, $\mu_j$ and $\eta_{ij}$ are   
\begin{eqnarray}\label{coeff}
\mu_1 &=& \frac{\chi^{2} \xi^{2}}{2}-\frac{4\chi v \xi^{2}}{3\textcolor{black}{q_0}} - \chi \xi - \frac{v^{2} \xi^{2}}{8\textcolor{black}{q_0^2}}+\frac{4v\xi}{3\textcolor{black}{q_0}}+\frac{2\pi^{2}\textcolor{black}{c^2}\xi^{2}}{\textcolor{black}{q_0^2}} +\frac{1}{2},
\nonumber\\
\mu_2 &=& \frac{\chi^{2} \xi^{2}}{2} + \frac{4 \chi v \xi^{2}}{3\textcolor{black}{q_0}} - \chi \xi - \frac{v^{2} \xi^{2}}{8\textcolor{black}{q_0^2}} - \frac{4 v \xi}{3\textcolor{black}{q_0}} + \frac{\pi^{2}\textcolor{black}{c^2} \xi^{2}}{2\textcolor{black}{q_0^2}} + \frac{1}{2},
\nonumber\\
\nu_1 &=& \frac{\chi^{2}}{2} + \frac{4 \chi v}{3\textcolor{black}{q_0}} - \frac{v^{2}}{8\textcolor{black}{q_0^2}} + \frac{\pi^{2}\textcolor{black}{c^2}}{2\textcolor{black}{q_0^2}},
\nonumber\\
\nu_2 &=& \frac{\chi^{2}}{2} - \frac{4 \chi v}{3\textcolor{black}{q_0}} - \frac{v^{2}}{8\textcolor{black}{q_0^2}} + \frac{2 \pi^{2}\textcolor{black}{c^2}}{\textcolor{black}{q_0^2}},
\nonumber\\
\eta_{12} &=&\chi^{2} \xi + \frac{8 \chi v \xi}{3\textcolor{black}{q_0}} - \chi - \frac{v^{2} \xi}{4\textcolor{black}{q_0^2}} - \frac{4 v}{3\textcolor{black}{q_0}} +\frac{\pi^{2}\textcolor{black}{c^2} \xi}{\textcolor{black}{q_0^2}},
\nonumber\\
\eta_{21} &=& \chi^{2} \xi - \frac{8 \chi v \xi}{3\textcolor{black}{q_0}} - \chi - \frac{v^{2} \xi}{4\textcolor{black}{q_0^2}} + \frac{4 v}{3\textcolor{black}{q_0}} + \frac{4 \pi^{2}\textcolor{black}{c^2} \xi}{\textcolor{black}{q_0^2}}.
\nonumber\\
\end{eqnarray}
The Hamiltonian $\hat{\mathcal{H}}_3$ reduces to its diagonal form whenever $\eta_{12}$ and $\eta_{21}$ vanish. 
This condition can be achieved by choosing $\chi$ and $\xi$ such that they satisfy
\begin{equation} \label{xi_chi}
\chi_{\pm}=\frac{9\pi^2\textcolor{black}{c^2}\pm\sqrt{\Gamma}}{16\textcolor{black}{q_0}v},
\qquad
\xi_{\pm}=\pm\frac{8\textcolor{black}{q_0}v}{\sqrt{\Gamma}},
\end{equation}
with $\Gamma=(8v^2+\pi^2\textcolor{black}{c^2})(81\pi^2\textcolor{black}{c^2}-8v^2)$.
However, since $\chi$ and $\xi$ appear in the definition of $\hat{R}_j$ in ~(\ref{RT}), it is imperative that $\Gamma\geq 0$ for $\hat{R}_j$ to remain as an unitary operator. This means that $v$ must satisfy the inequality $-\frac{9\pi}{2\sqrt{2}}\leq v \leq \frac{9\pi}{2\sqrt{2}}$.
Therefore, $\hat{\mathcal{H}}_3$ in its diagonal form is
$\hat{\mathcal{H}}_3=\sum_{j=1}^2(\mu_j\hat{p}_j^2+\nu_j\hat{x}_j^2)$,
where the coefficients $\nu_i$ and $\mu_j$ have been simplified to
\begin{equation}\label{muandnu}
\begin{split}
\mu_1=&\textcolor{black}{+}\frac{9 \pi ^2\textcolor{black}{c^2}}{4 \sqrt{\Gamma }}\textcolor{black}{-}\frac{16 v^2}{3 \sqrt{\Gamma }}+\frac{1}{4},\\
\mu_2=&\textcolor{black}{+}\frac{9 \pi ^2\textcolor{black}{c^2}}{4 \sqrt{\Gamma }}\textcolor{black}{+}\frac{16 v^2}{3 \sqrt{\Gamma }}+\frac{1}{4},\\
\nu_1=&\textcolor{black}{-}\frac{\sqrt{\Gamma }}{12\textcolor{black}{q_0^2}}+\frac{\Gamma }{256 v^2\textcolor{black}{q_0^2}}\textcolor{black}{-}\frac{9\textcolor{black}{c^2} \pi ^2 \sqrt{\Gamma }}{256 v^2\textcolor{black}{q_0^2}}, \\
\nu_2=&\textcolor{black}{+}\frac{\sqrt{\Gamma }}{12\textcolor{black}{q_0^2}}+\frac{\Gamma }{256 v^2\textcolor{black}{q_0^2}}\textcolor{black}{-}\frac{9 \pi ^2 \textcolor{black}{c^2}\sqrt{\Gamma }}{256 v^2\textcolor{black}{q_0^2}}.
\end{split}
\end{equation}
The Hamiltonian $\hat{\mathcal{H}}_3$ represents, in its diagonal form, two uncoupled quantum harmonic oscillators, and its eigenvalues are well known; we write them below Eq.~(\ref{twomodeversion}) of the main text. The above results correspond to the plus sign in~(\ref{xi_chi}), and similar expressions occur for the minus sign. We want to emphasize that this diagonalization procedure is original and nontrivial. 

%-------------------------------------------------------

%

\end{document}